\documentclass[runningheads]{llncs}
\usepackage[T1]{fontenc}
\usepackage{graphicx}
\usepackage[pagebackref=true,breaklinks=true,letterpaper=true,colorlinks,bookmarks=false]{hyperref}
\usepackage{color}

\usepackage{comment}
\usepackage{bm}
\usepackage{caption}
\usepackage{subcaption}
\usepackage{booktabs}
\usepackage{bold-extra}
\usepackage{xspace}
\usepackage[table,xcdraw]{xcolor}
\usepackage{dsfont}
\usepackage{colortbl}
\usepackage[frozencache]{minted}
\usepackage{textcomp}
\usepackage{multirow}
\usepackage{algorithm}
\usepackage{algorithmic}
\usepackage{amsmath}
\usepackage{makecell}
\usepackage{amssymb}
\usepackage[figuresright]{rotating}
\usepackage[misc]{ifsym}

\makeatother

\begin{document}
\def\thefootnote{*}\footnotetext{These authors contributed equally to this work.}

\title{Topology-Preserving Automatic Labeling of Coronary Arteries via Anatomy-aware Connection Classifier}

\titlerunning{Topology-Preserving Artery Labeling}
%
\author{Zhixing Zhang\thefootnote{} \inst{*1} \and
Ziwei Zhao\thefootnote{} \inst{*1,6} \and
Dong Wang\thefootnote{} \inst{2} \and 
Shishuang Zhao \inst{3} \and 
Yuhang Liu \inst{3} \and 
Jia Liu \inst{4} \and 
Liwei Wang \inst{2,5(\textrm{\Letter})}}
%


\authorrunning{Z. Zhang et al.}
%
\institute{Center for Data Science, Peking University, Beijing, China \\
\email{zhangzhixing@stu.pku.edu.cn}
\and
National Key Laboratory of General Artificial Intelligence, School of Intelligence Science and Technology,
Peking University, Beijing, China   \\
\email{wanglw@pku.edu.cn}
\and
Yizhun Medical AI Co., Ltd, Beijing, China  \\
\and
Peking University First Hospital, Beijing, China  \\
\and
Center for Machine Learning Research,  Peking University, Beijing, China \\
\and{Pazhou Lab, Guangzhou, China}
\\ 
}
\maketitle              
\begin{abstract}

Automatic labeling of coronary arteries is an essential task in the practical diagnosis process of cardiovascular diseases. For experienced radiologists, the anatomically predetermined connections are important for labeling the artery segments accurately, while this prior knowledge is barely explored in previous studies. In this paper, we present a new framework called TopoLab which incorporates the anatomical connections into the network design explicitly. Specifically, the strategies of intra-segment feature aggregation and inter-segment feature interaction are introduced for hierarchical segment feature extraction. Moreover, we propose the anatomy-aware connection classifier to enable classification for each connected segment pair, which effectively exploits the prior topology among the arteries with different categories. To validate the effectiveness of our method, we contribute high-quality annotations of artery labeling to the public orCaScore dataset. The experimental results on both the orCaScore dataset and an in-house dataset show that our TopoLab has achieved state-of-the-art performance.

\end{abstract}

\section{Introduction}

\begin{figure}[t]
    \centering
    \includegraphics[width=1.0\textwidth]{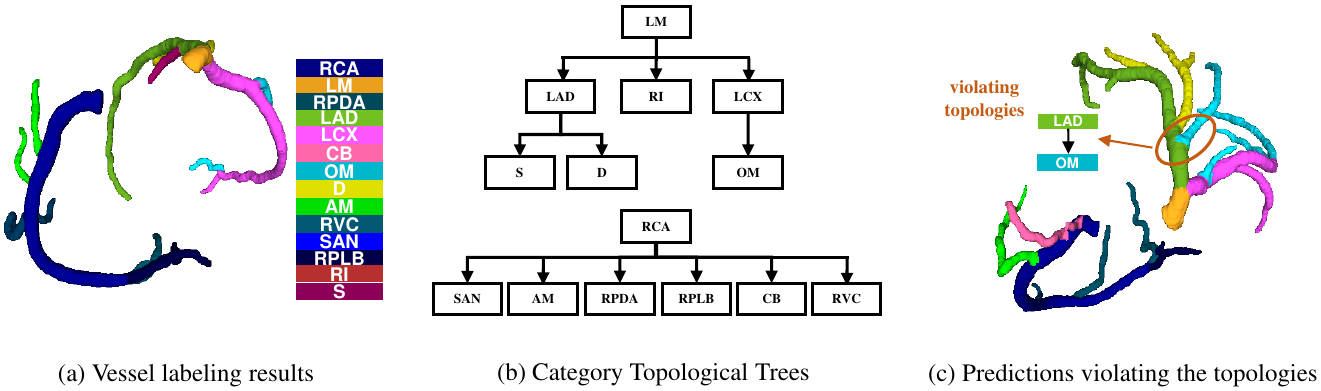}
    \caption{(a) The task of coronary artery labeling aims to assign each vessel segment an anatomical category. (b) The anatomically predetermined connections among different categories of coronary arteries form a tree structure -- \emph{category topological trees}.  (c) shows an example containing obvious mistakes violating the prior topology made by previous methods.}
    \label{fig:intro}
\end{figure}

Coronary Computerized Tomography Angiography (CCTA) is a commonly used non-invasive approach for the diagnosis of potential coronary artery diseases~\cite{mowatt200864}. In clinical practice, accurate labeling of coronary artery segments (see Figure~\ref{fig:intro}(a)) is a crucial step toward the subsequent diagnosis and analysis of the image. However, the vast variability of coronary artery anatomy across individuals makes it challenging to achieve precise and automatic labeling.

Previous studies on labeling coronary arteries using deep learning-based methods~\cite{wu2019automated,yang2020cpr,yao2023tag,zhang2021corlab} have shown promising results by introducing graph convolutional networks and point cloud analysis. However, these approaches have overlooked the essential prior knowledge that \textbf{different categories of coronary arteries have anatomically predetermined connections}~\cite{dodge1988intrathoracic}. For instance, LAD, LCX,  and RI originate from LM, while S and D arise from LAD. All of the anatomical connections form a tree structure with prior topology, as shown in Figure~\ref{fig:intro}(b). We call this structure \emph{category topological trees}. Due to lacking the utilization of the \emph{category topological trees}, existing methods~\cite{wu2019automated,yang2020cpr,yao2023tag,zhang2021corlab} often make some obvious mistakes that violate the prior topology as illustrated in Figure~\ref{fig:intro}(c). We argue that incorporating the \emph{category topological trees} into the network explicitly is the key to improving automatic labeling performance, especially in reducing topology-violation labeling errors.

In this paper, we propose a novel framework called TopoLab to perform topology-preserving automatic labeling of coronary arteries. Our model mainly contains two components: the hierarchical feature extraction module and the anatomy-aware connection classifier. The hierarchical feature extraction module introduces the segment query to achieve intra-segment feature aggregation via Transformer~\cite{vaswani2017attention} and relies on graph convolutional network~\cite{kipf2016semi} to establish inter-segment feature interactions. Moreover, to incorporate the \emph{category topological trees} into the network explicitly, we further propose the anatomy-aware connection classifier (AC-Classifier). Unlike previous methods that classify each segment independently, AC-Classifier performs classification for every connected segment pair. 
Specifically, all of the connections derived from the \emph{category topological trees} are used to construct the ground truth connection templates, and each connected segment pair is categorized into one of these templates. Since the connection templates inherently conform to the topology, AC-Classifier has effectively prioritized the anatomically predetermined connections and the network is enabled to preserve the topology by design. 

To the best of our knowledge, there is currently no publicly available dataset with annotations for artery labeling. In this work, we contribute high-quality annotations to the orCaScore dataset~\cite{wolterink2016evaluation}.
The experimental results on the public dataset orCaScore and an in-house dataset have demonstrated that our TopoLab outperforms previous state-of-the-art methods, especially in the topology-related metrics.

Our contributions can be summarized as follows. (1) We are the first to incorporate the \emph{category topological trees} into the deep learning models for automatic labeling of coronary arteries by introducing AC-Classifier. 
(2) We propose a novel hierarchical feature extraction module to achieve intra- and inter-segment feature aggregations.
(3) Our approach achieves state-of-the-art performance on both public and in-house datasets. 
(4) We provide high-quality annotations of artery labeling for the public dataset orCaScore, which is available at \url{https://github.com/zutsusemi/MICCAI2023-TopoLab-Labels/}.

\section{Related Work}
Traditional methods~\cite{cao2017automatic,yang2011automatic} for automatic labeling of coronary arteries usually align the extracted artery trees with a 3D
coronary artery tree model which provides the anatomical connections as prior knowledge. 
However, these works rely heavily on logical rules, which can not always capture the complexities of the anatomical structure.
To overcome the limitations, deep learning has been introduced in this area~\cite{wu2019automated,yang2020cpr,yao2023tag,zhang2021corlab} with its great success in medical imaging~\cite{ronneberger2015u,isensee2021nnu,shen2017deep,shit2021cldice,wang2022pointscatter}. 
For instance, TreeLab-Net~\cite{wu2019automated} uses bidirectional tree-structural LSTM~\cite{graves2005framewise} to model the coronary artery trees, while CPR-GCN~\cite{yang2020cpr} constructs a vessel graph by treating each segment as a node and leverages GCN~\cite{kipf2016semi} to aggregate segment features.  
CorLab-Net~\cite{zhang2021corlab} regards spatial and anatomical dependencies as the explicit guidance for artery labeling based on point cloud networks~\cite{qi2017pointnet++}. Nevertheless, all of these deep learning based methods ignore the important priority about the predetermined anatomical structure -- \emph{category topological trees}. In this paper, we aim to incorporate this priority into the network design explicitly for developing the topology-preserving models.

\section{Methodology}

\subsection{Overview}
\label{sec ove}
We start by extracting centerlines from the vessel segmentation annotations in a CCTA image, using a traditional 3D thinning algorithm~\cite{lee1994building}. Next, we use the minimum spanning tree algorithm~\cite{graham1985history} to construct two coronary artery trees, one for the left domain (LD) and one for the right domain (RD). Following the branch bifurcation rules in~\cite{yang2020cpr} (details can be found in supplementary materials), the coronary artery trees are split into several segments, denoted by $\mathcal{S}=\{S_i\}_{i=1}^N$, where $N$ is the number of segments, $S_i \in \mathbb{R}^{L_i \times 3}$ denotes the $i$-th segment comprised of the 3D positions of $L_i$ centerline points. Our model takes as input the vessel segments $\mathcal{S}$ and their connections $\mathcal{C}=\{(i_1, j_1),(i_2, j_2),...,(i_{N_c}, j_{N_c})\}$, where $(i_k, j_k)$ indicates that the $i_k$-th segment is connected with the $j_k$-th segment. The objective is to predict the classes of the vessel segments. 

\begin{figure}[t]
    \centering
    \includegraphics[width=1.0\textwidth]{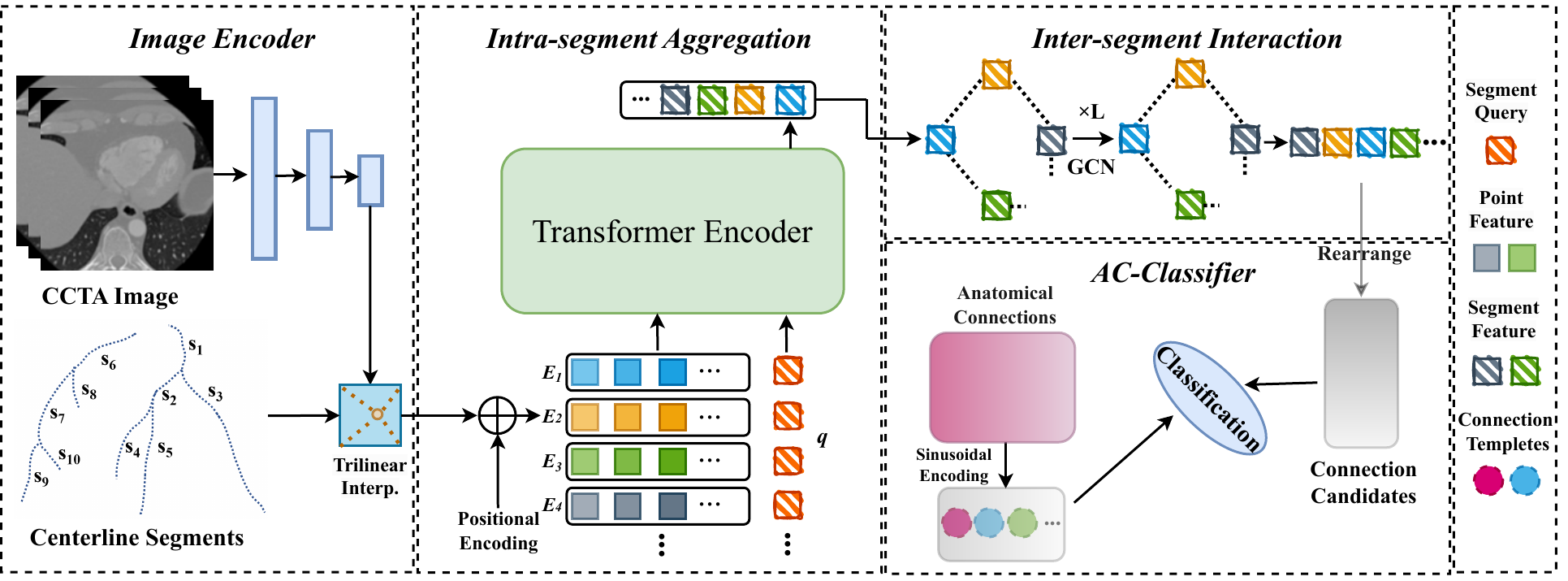}
    \caption{The pipeline of our proposed TopoLab.}
    \label{fig:model}
\end{figure}

As illustrated in Figure~\ref{fig:model}, we design a novel framework named TopoLab for the automatic labeling of coronary arteries. In Sec.~\ref{Sec: feat extract}, we will introduce the hierarchical feature extraction module comprised of intra-segment feature aggregation and inter-segment feature interaction. In Sec.~\ref{Sec: classifier}, we will elaborate on the details of AC-Classifier which exploits the \emph{category topological trees} effectively.

\subsection{Hierarchical Feature Extraction}
\label{Sec: feat extract}
We first feed the CCTA image $X \in \mathbb{R}^{H \times W \times D}$ into an image encoder (\textit{e.g.} U-Net \cite{cciccek20163d}) to obtain the downscaled feature map $F \in \mathbb{R}^{H/4 \times W/4 \times D/4 \times C}$, where $C$ is the channel size. 

For each segment $S_i$, trilinear interpolation in the downsampled feature map $F$ is adopted for the centerline point $v \in S_i$ to obtain the corresponding point features $ f(v) = \text{Tri}(v, F) \in \mathbb{R}^C.$ The segment features are denoted as $\{f(v) | v \in S_i\}$, simplified as $E_i \in \mathbb{R}^{L_i \times C}$.

\noindent \textbf{Intra-Segment Feature Aggregation.}
To extract the segment-level features for labeling, aggregation of sequential point features belonging to the same segment is required. Previous studies~\cite{wu2019automated,yang2020cpr} employ Bidirectional LSTM~\cite{graves2005framewise} to summarize the tubular sequential features. However, due to the weak representative ability of LSTM for the vessel segments with huge length variability, the performance of these methods is limited. 

We introduce Transformer~\cite{vaswani2017attention} as our intra-segment feature aggregator for its strong capability to model the relationships among the sequences with varying lengths. Concretely, we set a learnable embedding $q \in \mathbb{R}^C$ called segment query to aggregate intra-segment point features. The segment query is concatenated with the segment features $E_i$ to obtain the new tensor $ \tilde{E}_i \in \mathbb{R}^{(L_i+1) \times C}$. Following~\cite{dosovitskiy2020image}, we feed $ \tilde{E}_i$ augmented by the learnable 3D positional encodings into a Transformer encoder containing several standard sub-blocks. Each standard sub-block has the same architecture as in~\cite{vaswani2017attention}, which consists of multi-head attention, feed-forward network, layer normalization and ReLU activation. The state of the segment query $q$ at the output of the Transformer encoder serves as the aggregated segment representation $\hat{E}_i \in \mathbb{R}^C$.

\noindent \textbf{Inter-Segment Feature Interaction.}
The branching structure of coronary arteries is inherently graph-like, with each segment serving as a node and the connections between segments serving as edges. Thus, we leverage graph convolutional network~\cite{kipf2016semi} (GCN) to capture the interactions among different segments.

Specifically, let $\hat{E} \in \mathbb{R}^{N \times C} $ denotes the aggregated segment features where the $i$-th item of $\hat{E}$ is $\hat{E}_i$, and $A \in \mathbb{R}^{N \times N} $ is the adjacency matrix for the vessel segment graph derived from the segment connections $\mathcal{C}$. The 
process of GCN layers is as follows:
\begin{equation}
    \hat{E}_{l+1} = \sigma(A\hat{E}_lW_l),
\end{equation}
where $\sigma$ is the ReLU activation, $W_l \in \mathbb{R}^{C \times C}$ is the learnable parameters for the $l$-th GCN layer, the input for the first layer is $\hat{E}_0=\hat{E}$.

Finally, we fuse the input segment features $\hat{E}$ and the output of the final GCN layer $\hat{E}_{f}$ to obtain $\overline{E}=[\hat{E}_f, \hat{E}]W_f \in \mathbb{R}^{N\times C}$ with the parameters $W_f\in \mathbb{R}^{(C+C)\times C}$. The enhanced segment features $\{\overline{E}_1, \overline{E}_2, ..., \overline{E}_N\}$ are forwarded to the classifier for segment labeling, where $\overline{E}_i \in \mathbb{R}^C$ is the $i$-th item of $\overline{E}$.

\subsection{Anatomy-aware Connection Classifier}
\label{Sec: classifier}
The direct approach for labeling the coronary arteries is to use a linear layer to classify each segment independently as in previous methods. 
To incorporate the \emph{category topological trees} into the classifier design, we propose to conduct the classification task for every connected segment pair. 

We begin by defining the ground truth segment connections which are composed of the topology-conforming connections derived from the \emph{category topological trees} (like LM$\rightarrow$LAD, LCX$\rightarrow$OM, \textit{etc.}). Note that the self-connections (\textit{e.g.}  RCA$\rightarrow$RCA) are also considered as the ground truth connections.
The ground truth segment connections have the corresponding template embeddings for classification, which are represented by $G \in \mathbb{R}^{N_g \times 2C}$, where $N_g$ is the number of ground truth segment connections. Denote $g_i=\text{Concate}(\text{Enc}(x), \text{Enc}(y)) \in \mathbb{R}^{2C}$ as the $i$-th item of $G$, where $x$ and $y$ stand for the segment classes indexed by the $i$-th ground truth connection, Enc denotes the sinusoidal encoding as in \cite{vaswani2017attention}. The connection templates $G$ are used to enable classification for the connected segment pairs.

Then, given the segment features $\{\overline{E}_i\}_{i=1}^N$, and all of the connected segment pairs $\mathcal{C}=\{(i_1, j_1),(i_2, j_2),...,(i_{N_c}, j_{N_c})\}$, the connection features $P \in \mathbb{R}^{N_c \times 2C}$ are obtained by rearrangement of the segment features, where the $k$-th item of $P$ is $\text{Concate}(\overline{E}_{i_k}, \overline{E}_{j_k}) \in \mathbb{R}^{2C}$. 
We use an MLP layer to further fuse the features $\hat{P} = \text{MLP}(P) \in \mathbb{R}^{N_c \times 2C}$.

\noindent \textbf{Training Loss.} The loss function can be written as: 

\begin{equation}
    \mathcal{L} = \sum_i^{N_c} \mathcal{L}_{\text{cls}}(\hat{p}_i, y_i),
\end{equation}

\begin{equation}
\label{equ: loss}
    \mathcal{L}_{\text{cls}} = -\text{log}\frac{\text{exp}(\text{sim}(\hat{p}_i, g_{y_i})/\tau)}{\sum_{j=1}^{N_g}\text{exp}(\text{sim}(\hat{p}_i, g_{j})/\tau)},
\end{equation}
where $y_i$ is the ground truth of the $i$-th segment connection, $\hat{p}_i$ denotes the $i$-th item of $\hat{P}$. \text{sim} in Eq.~\ref{equ: loss} stand for the cosine similarity, $\text{sim}(x, y) = \frac{x^T y}{||x||_2||y||_2}$, and the temperature $\tau$ is a hyperparameter which is set as $0.05$ by default. 

\noindent \textbf{Inference.} 
During the inference stage, for each segment, we first select the connection with the largest confidence score among all segment connections that have covered the given segment. And then the corresponding category indexed by the selected connection serves as the prediction of the specific segment.

\section{Experiments}

\subsection{Setup} \label{sec:setup}

\textbf{Datasets.} We train and evaluate our method on two datasets. The \textbf{orCaScore}~\cite{wolterink2016evaluation} MICCAI 2014 Challenge contains 72 contrast-enhanced CTA images and non-contrast enhanced CT scans. As the original dataset only contains the labels of calcifications, we have annotated vessel segmentation and anatomical categories for coronary arteries with experienced radiologists. Considering the small amount of data, we randomly split the dataset into five folds to perform cross-validation. The mean values of cross-validation results are reported. We also collect an \textbf{In-house Dataset} containing 1200 CTA scans which have been annotated by at least two experts. The dataset is collected in compliance with the terms of the licensing agreement and ethical certification. We randomly split the dataset into train, validation and test set with 800, 200 and 200 scans respectively. The annotations for both datasets adhere to the same standard, which includes 14 classes of coronary artery segments (see Figure~\ref{fig:intro}(b)). It is a challenging task, surpassing the scope of studies like TreeLab-Net\cite{wu2019automated} and CPR-GCN\cite{yang2020cpr}, which only consider 10 and 11 categories, respectively.

\noindent \textbf{Evaluation Metrics.} Following previous methods~\cite{yang2020cpr,wu2019automated,zhang2021corlab}, we adopt the mean metrics of all categories of the segments including recall, precision and F1. Note that the mean metric is the weighted average based on the number of segments of different categories. To further evaluate the topological accuracy of connected segments, we propose two new metrics: viola and $\text{viola}^\text{c}$, which reflect the segment-level topological accuracy and the case-level topological accuracy, respectively. Specifically, $\text{viola}$ is calculated as the ratio of the number of connections violating the topology to the total number of connections, while $\text{viola}^\text{c}$ is calculated as the ratio of the number of test cases containing any topology-violating connection to the total number of test cases. 

\subsection{Implementation Details}
3D ResUNet~\cite{zhang2018road} is employed as the image encoder for feature extraction with channel dimension $C=64$. The transformer encoder has 3 standard blocks and the number of graph convolution layers is set to 4. We train the network using AdamW optimizer~\cite{loshchilov2017decoupled} with a base learning rate of 5e-4 and the cosine learning rate schedule during the training stage. The batch size is set to 4. All networks are implemented by Pytorch~\cite{adam19torch} and trained on four NVIDIA GeForce RTX 3090 GPUs. We train our model on the orCaScore dataset for 3.5k iterations and in-house dataset for 12.5k iterations.

\subsection{Comparison with Other Methods}
\noindent\textbf{Quantitative Results.} We compare TopoLab with other deep learning based approaches including TaG-Net~\cite{yao2023tag},
CPR-GCN~\cite{yang2020cpr}, TreeLab-Net~\cite{wu2019automated}, and CorLab-Net~\cite{zhang2021corlab} which are implemented by ourselves with the same training configurations for a fair comparison. Note that for the point cloud-based methods~\cite{zhang2021corlab,yao2023tag}, we use intra-segment voting to transform the point-level predictions into segment-level predictions. From the results on the orCaScore dataset in Table~\ref{tab: sota1} and the in-house dataset in Table~\ref{tab: sota2}, we can conclude that the proposed TopoLab outperforms all existing methods by a large margin. The performance gains are more significant in the topology-related metrics, which demonstrates the effectiveness of the utilization of prior knowledge. More detailed results on each category of coronary arteries can be found in the supplementary materials.

\begin{figure}[t]
    \centering
    \includegraphics[width = 0.9\textwidth]{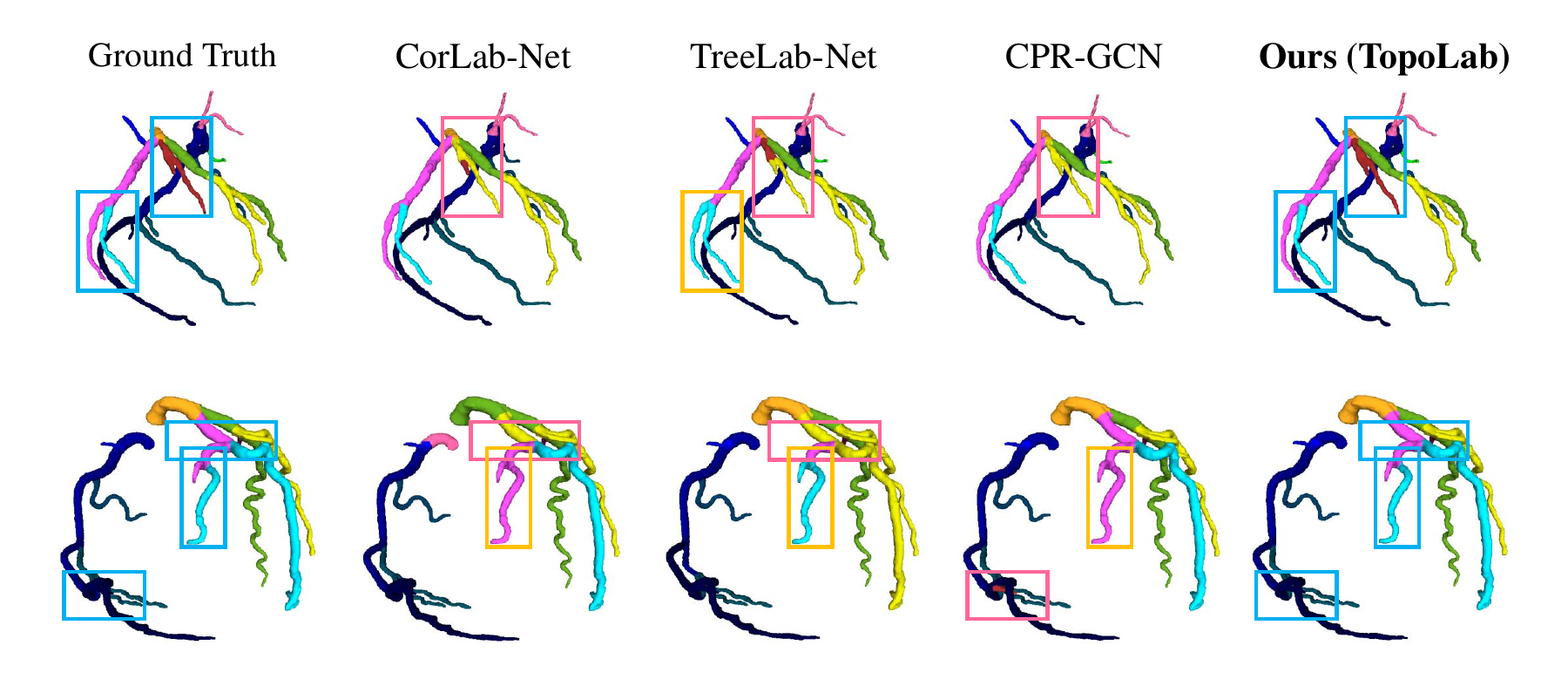}
    \caption{Visual comparison between our TopoLab and other methods. Errors violating topology are highlighted in red boxes, while errors that do not violate topology are marked in orange boxes. The corresponding positions in the ground truth and TopoLab predictions are indicated by blue boxes. }
    \label{fig: visual}
\end{figure}

\noindent\textbf{Qualitative Results.} In Figure~\ref{fig: visual}, we present a qualitative comparison of TopoLab with other methods. Consider the first case as an example, where our approach successfully avoids topology-violating errors, whereas other methods incorrectly classify RI as D, leading to the RI$\rightarrow$D, D$\rightarrow$RI, or LM$\rightarrow$D connections that violate topology. These visualizations demonstrate the effectiveness of the usage of \emph{category topological trees}. More qualitative results can be found in supplementary materials.

\begin{table}[t]
\caption{Comparison with other methods on orCaScore dataset(\%). Each entry of the table shows the average value of 5 folds with the standard deviation in subscript. "Recall, precision, and F1" of each fold are the weighted averages of the 14 vessel categories.}
\centering
\label{tab: sota1}
\setlength{\tabcolsep}{4pt}{
\scalebox{0.9}[0.9]{
\begin{tabular}{lccccc}
\toprule[1.0pt]
Method   & Recall & Precison & F1   & Viola  & $\text{Viola}^\text{c}$      \\ \midrule
TaG-Net~\cite{yao2023tag}& $82.29_{2.07}$ & $83.41_{2.28}$ & $82.14_{2.32}$          & $10.87_{2.49}$         & $87.62_{4.85}$       \\
CorLab-Net~\cite{zhang2021corlab} & 
$ 82.09_{1.03}$&$83.83_{1.53}$ & $82.15_{1.18}$      & $9.01_{1.09}$    &  $75.05_{10.01}$      \\
TreeLab-Net~\cite{wu2019automated} & $83.35_{2.80}$ & $84.90_{2.11}$ & $83.12_{2.96}$          & $3.75_{1.08}$          & $55.52_{17.26} $        \\

CPR-GCN~\cite{yang2020cpr}  & $82.88_{1.44}$ & $83.61_{1.59}$ & $82.72_{1.55}$          & $4.94_{1.49}$         & $53.91_{9.53}$        \\

\textbf{TopoLab(Ours)} & $\textbf{87.13}_{1.03}$ & $\textbf{88.31}_{1.60}$&$\textbf{87.23}_{1.29}$ & $\textbf{1.52}_{0.73}$ & $\textbf{22.29}_{8.30}$\\
\bottomrule[1.0pt]
\end{tabular}}}
\end{table}

\begin{table}[t]
\caption{Comparison with other methods on in-house dataset(\%). Each entry of the table shows the average value of 5 trials with the standard deviation in subscript. "Recall, precision, and F1" of each trial are the weighted averages of the 14 vessel categories.}
\centering
\label{tab: sota2}
\setlength\tabcolsep{4pt} 
\scalebox{0.9}[0.9]{
\begin{tabular}{lccccc}
\toprule[1.0pt]
Method &   Recall & Precision & F1 & Viola  & $\text{Viola}^\text{c}$  \\ \midrule
TaG-Net~\cite{yao2023tag}& $88.26_{0.30}$ & $88.47_{0.34}$ & $88.23_{0.34}$        & $6.46_{0.28}$         & $65.90_{1.98}$         \\
CorLab-Net~\cite{zhang2021corlab} & $88.85_{0.35}$ & $88.96_{0.34}$ & $88.83_{0.35}$        & $4.75_{0.31}$         & $57.10_{1.46}$          \\
TreeLab-Net~\cite{wu2019automated} & $88.58_{0.43}$ & $88.56_{0.44}$ & $88.52_{0.44}$        & $2.34_{0.09}$         & $32.70_{0.51}$              \\
CPR-GCN~\cite{yang2020cpr}  & $90.92_{0.15}$ & $90.84_{0.13}$ &$90.84_{0.14}$        &$ 2.89_{0.17}$         & $38.70_{3.93}$    \\
\textbf{TopoLab(Ours)} & $\textbf{92.23}_{0.23}$&$\textbf{92.21}_{0.25}$     & $\textbf{92.19}_{0.23}$ & $\textbf{0.77}_{0.13}$ & $\textbf{9.40}_{1.69} $    \\ 
\bottomrule[1.0pt]
\end{tabular}
}
\end{table}

\subsection{Ablation Study}

In this subsection, we explore the effectiveness of different components in TopoLab on orCaScore dataset, as shown in Table~\ref{tab: ablation}.

\noindent{\textbf{Intra-segment Feature Aggregation (IFA).}} Transformer~\cite{vaswani2017attention} is leveraged to achieve the intra-segment feature aggregation in our model. To validate its benefits, we replace it with the Bi-LSTM used in CPR-GCN. From the results in the first line of Table~\ref{tab: ablation}, our method surpasses Bi-LSTM by 7.94\% in F1 score and 26.57\% in $\text{Viola}^\text{c}$.

\noindent{\textbf{Inter-segment Feature Interaction (IFI).}} We use GCN to establish inter-segment feature interactions due to the natural graph structure of coronary arteries. When directly removing this module, the performance drops by 1.57\% in F1 score and 19.14\% in $\text{Viola}^\text{c}$ as illustrated in the second line of Table~\ref{tab: ablation}.

\noindent{\textbf{Anatomy-aware Connection Classifier (ACC).}} 
AC-Classifier which exploits the prior knowledge from \emph{category topological trees} is adopted to classify each connected segment pair. We replace it with a commonly used linear layer to enable classification for single segments as in previous methods, and the performance drops by 1.33\% in $\text{Viola}$ and 15.42\% in $\text{Viola}^\text{c}$, which effectively demonstrates the superiority of the proposed method.

\begin{table}[t]
\caption{Ablation Study on orCaScore Dataset(\%). Each entry of the table shows the average value of 5 folds. "Recall, precision, and F1" of each fold are the weighted averages of the 14 vessel categories.}
\centering
\label{tab: ablation}
\setlength\tabcolsep{4pt} 
\scalebox{0.9}[0.9]{
\begin{tabular}{lccccc}
\toprule[1.0pt]
Method   &  Recall &  Precision & F1   & Viola  & $\text{Viola}^\text{c}$  \\
\midrule
w/o IFA &$79.61_{3.38}$& $81.47_{2.71}$ & $79.29_{3.35}$  & $4.00_{1.82}$  & $48.86_{20.39}$                      \\
w/o IFI &  $85.79_{2.61}$&   $86.74_{2.77}$ & $85.66_{2.83}$   & $2.98_{1.54}$ &  $41.43_{15.55}$                  \\
w/o ACC   &$86.88_{1.39}$&  $88.15_{1.74}$&  $86.95_{1.57}$& $2.85_{0.85}$&  $37.71_{10.26}$    \\
\textbf{TopoLab} & $\textbf{87.13}_{1.03}$ & $\textbf{88.31}_{1.60}$&$\textbf{87.23}_{1.29}$ & $\textbf{1.52}_{0.73}$ & $\textbf{22.29}_{8.30}$\\
\bottomrule[1.0pt]
\end{tabular}}
\end{table}

\section{Conclusion}
In this study, we review the essential task of coronary artery labeling and exploit the prior knowledge of the predetermined anatomical connections. The proposed strategies of intra- and inter-segment feature aggregation guarantee effective feature extraction, while the AC-Classifier preserves the clinical logic in the network design. The extensive experiments on orCaScore dataset and in-house dataset reveal that the proposed TopoLab has achieved new state-of-the-art performance. We hope our paper could encourage the community to explore the use of clinical priority to facilitate the design of more effective algorithms.

\paragraph{\emph{\textbf{Acknowledgements.}}} We would like to thank Dr. Nianxi Liao for valuable discussions. This work is supported by National Key R\&D Program of China (2022ZD0114900) and National Science Foundation of China (NSFC62276005). 

%
%
%
\bibliographystyle{splncs04}
\bibliography{mybibliography}

\begin{thebibliography}{10}
\providecommand{\url}[1]{\texttt{#1}}
\providecommand{\urlprefix}{URL }
\providecommand{\doi}[1]{https://doi.org/#1}

\bibitem{cao2017automatic}
Cao, Q., Broersen, A., de~Graaf, M.A., Kitslaar, P.H., Yang, G., Scholte, A.J.,
  Lelieveldt, B.P., Reiber, J.H., Dijkstra, J.: Automatic identification of
  coronary tree anatomy in coronary computed tomography angiography. The
  international journal of cardiovascular imaging  \textbf{33},  1809--1819
  (2017)

\bibitem{cciccek20163d}
{\c{C}}i{\c{c}}ek, {\"O}., Abdulkadir, A., Lienkamp, S.S., Brox, T.,
  Ronneberger, O.: 3d u-net: learning dense volumetric segmentation from sparse
  annotation. In: International conference on medical image computing and
  computer-assisted intervention. pp. 424--432. Springer (2016)

\bibitem{dodge1988intrathoracic}
Dodge~Jr, J.T., Brown, B.G., Bolson, E.L., Dodge, H.T.: Intrathoracic spatial
  location of specified coronary segments on the normal human heart.
  applications in quantitative arteriography, assessment of regional risk and
  contraction, and anatomic display. Circulation  \textbf{78}(5),  1167--1180
  (1988)

\bibitem{dosovitskiy2020image}
Dosovitskiy, A., Beyer, L., Kolesnikov, A., Weissenborn, D., Zhai, X.,
  Unterthiner, T., Dehghani, M., Minderer, M., Heigold, G., Gelly, S., et~al.:
  An image is worth 16x16 words: Transformers for image recognition at scale.
  arXiv preprint arXiv:2010.11929  (2020)

\bibitem{graham1985history}
Graham, R.L., Hell, P.: On the history of the minimum spanning tree problem.
  Annals of the History of Computing  \textbf{7}(1),  43--57 (1985)

\bibitem{graves2005framewise}
Graves, A., Schmidhuber, J.: Framewise phoneme classification with
  bidirectional lstm and other neural network architectures. Neural networks
  \textbf{18}(5-6),  602--610 (2005)

\bibitem{isensee2021nnu}
Isensee, F., Jaeger, P.F., Kohl, S.A., Petersen, J., Maier-Hein, K.H.: nnu-net:
  a self-configuring method for deep learning-based biomedical image
  segmentation. Nature methods  \textbf{18}(2),  203--211 (2021)

\bibitem{kipf2016semi}
Kipf, T.N., Welling, M.: Semi-supervised classification with graph
  convolutional networks. arXiv preprint arXiv:1609.02907  (2016)

\bibitem{lee1994building}
Lee, T.C., Kashyap, R.L., Chu, C.N.: Building skeleton models via 3-d medial
  surface axis thinning algorithms. CVGIP: graphical models and image
  processing  \textbf{56}(6),  462--478 (1994)

\bibitem{loshchilov2017decoupled}
Loshchilov, I., Hutter, F.: Decoupled weight decay regularization. arXiv
  preprint arXiv:1711.05101  (2017)

\bibitem{mowatt200864}
Mowatt, G., Cook, J.A., Hillis, G.S., Walker, S., Fraser, C., Jia, X., Waugh,
  N.: 64-slice computed tomography angiography in the diagnosis and assessment
  of coronary artery disease: systematic review and meta-analysis. Heart
  \textbf{94}(11),  1386--1393 (2008)

\bibitem{adam19torch}
Paszke, A., Gross, S., Massa, F., Lerer, A., Bradbury, J., Chanan, G., Killeen,
  T., Zeming~Lin, e.a.: Pytorch: An imperative style, high-performance deep
  learning library. In: Wallach, H.M., Larochelle, H., Beygelzimer, A.,
  d'Alch{\'{e}}{-}Buc, F., Fox, E.B., Garnett, R. (eds.) NeurIPS (2019)

\bibitem{qi2017pointnet++}
Qi, C.R., Yi, L., Su, H., Guibas, L.J.: Pointnet++: Deep hierarchical feature
  learning on point sets in a metric space. Advances in neural information
  processing systems  \textbf{30} (2017)

\bibitem{ronneberger2015u}
Ronneberger, O., Fischer, P., Brox, T.: U-net: Convolutional networks for
  biomedical image segmentation. In: Medical Image Computing and
  Computer-Assisted Intervention--MICCAI 2015: 18th International Conference,
  Munich, Germany, October 5-9, 2015, Proceedings, Part III 18. pp. 234--241.
  Springer (2015)

\bibitem{shen2017deep}
Shen, D., Wu, G., Suk, H.I.: Deep learning in medical image analysis. Annual
  review of biomedical engineering  \textbf{19},  221--248 (2017)

\bibitem{shit2021cldice}
Shit, S., Paetzold, J.C., Sekuboyina, A., Ezhov, I., Unger, A., Zhylka, A.,
  Pluim, J.P., Bauer, U., Menze, B.H.: cldice-a novel topology-preserving loss
  function for tubular structure segmentation. In: Proceedings of the IEEE/CVF
  Conference on Computer Vision and Pattern Recognition. pp. 16560--16569
  (2021)

\bibitem{vaswani2017attention}
Vaswani, A., Shazeer, N., Parmar, N., Uszkoreit, J., Jones, L., Gomez, A.N.,
  Kaiser, {\L}., Polosukhin, I.: Attention is all you need. Advances in neural
  information processing systems  \textbf{30} (2017)

\bibitem{wang2022pointscatter}
Wang, D., Zhang, Z., Zhao, Z., Liu, Y., Chen, Y., Wang, L.: Pointscatter: Point
  set representation for tubular structure extraction. In: Computer
  Vision--ECCV 2022: 17th European Conference, Tel Aviv, Israel, October
  23--27, 2022, Proceedings, Part XXI. pp. 366--383. Springer (2022)

\bibitem{wolterink2016evaluation}
Wolterink, J.M., Leiner, T., De~Vos, B.D., Coatrieux, J.L., Kelm, B.M., Kondo,
  S., Salgado, R.A., Shahzad, R., Shu, H., Snoeren, M., et~al.: An evaluation
  of automatic coronary artery calcium scoring methods with cardiac ct using
  the orcascore framework. Medical physics  \textbf{43}(5),  2361--2373 (2016)

\bibitem{wu2019automated}
Wu, D., Wang, X., Bai, J., Xu, X., Ouyang, B., Li, Y., Zhang, H., Song, Q.,
  Cao, K., Yin, Y.: Automated anatomical labeling of coronary arteries via
  bidirectional tree lstms. International journal of computer assisted
  radiology and surgery  \textbf{14},  271--280 (2019)

\bibitem{yang2011automatic}
Yang, G., Broersen, A., Petr, R., Kitslaar, P., de~Graaf, M.A., Bax, J.J.,
  Reiber, J.H., Dijkstra, J.: Automatic coronary artery tree labeling in
  coronary computed tomographic angiography datasets. In: 2011 Computing in
  Cardiology. pp. 109--112. IEEE (2011)

\bibitem{yang2020cpr}
Yang, H., Zhen, X., Chi, Y., Zhang, L., Hua, X.S.: Cpr-gcn: Conditional
  partial-residual graph convolutional network in automated anatomical labeling
  of coronary arteries. In: Proceedings of the IEEE/CVF conference on computer
  vision and pattern recognition. pp. 3803--3811 (2020)

\bibitem{yao2023tag}
Yao, L., Shi, F., Wang, S., Zhang, X., Xue, Z., Cao, X., Zhan, Y., Chen, L.,
  Chen, Y., Song, B., et~al.: Tag-net: Topology-aware graph network for
  centerline-based vessel labeling. IEEE Transactions on Medical Imaging
  (2023)

\bibitem{zhang2021corlab}
Zhang, X., Cui, Z., Feng, J., Song, Y., Wu, D., Shen, D.: Corlab-net:
  anatomical dependency-aware point-cloud learning for automatic labeling of
  coronary arteries. In: Machine Learning in Medical Imaging: 12th
  International Workshop, MLMI 2021, Held in Conjunction with MICCAI 2021,
  Strasbourg, France, September 27, 2021, Proceedings 12. pp. 576--585.
  Springer (2021)

\bibitem{zhang2018road}
Zhang, Z., Liu, Q., Wang, Y.: Road extraction by deep residual u-net. IEEE
  Geoscience and Remote Sensing Letters  \textbf{15}(5),  749--753 (2018)

\end{thebibliography}

\end{document}